%% file: paper.tex
\documentclass{fundam}
\publyear{25}
\papernumber{2}
\volume{194}
\issue{2}
\theDOI{10.46298/fi.12736}

\versionForARXIV

\input macros

\title{Peano Arithmetic and \muMALL}
\author{Matteo Manighetti\\
        Inria Saclay\\
        Saclay, France
    \and 
        Dale Miller\\
        Inria-Saclay and LIX, Institut Polytechnique de Paris\\
        Palaiseau, France}
\runninghead{M. Manighetti, D. Miller}{Peano Arithmetic and \mumall}
\date{\noindent Draft: \today}
\begin{document}
\maketitle

\begin{abstract}
Formal theories of arithmetic have traditionally been based on either
classical or intuitionistic logic, leading to the development of Peano
and Heyting arithmetic, respectively.  We propose to use $\mu$MALL as a
formal theory of arithmetic based on linear logic.  This formal system
is presented as a sequent calculus proof system that extends the
standard proof system for multiplicative-additive linear logic (MALL)
with the addition of the universal and existential
quantifiers (first-order quantifiers), term equality and non-equality,
and the least and greatest fixed point operators.  We first
demonstrate how functions defined using $\mu$MALL relational
specifications can be computed using a simple proof search algorithm.
By incorporating weakening and contraction into $\mu$MALL, we obtain
$\mu$LK$_p^+$, a natural candidate for a classical sequent calculus for
arithmetic.  While important proof theory results are still lacking
for $\mu$LK$_p^+$ (including cut-elimination and the completeness of
focusing), we prove that $\mu$LK$_p^+$ is consistent and that it contains
Peano arithmetic.  We also prove some conservativity results regarding
$\mu$LK$_p^+$ over $\mu$MALL.
\end{abstract}

\input intro.tex     % Section 1
\input terms.tex     % Section 2
\input polarized.tex % Section 3
\input compute.tex   % Section 4
\input ackermann.tex % Section 5
\input peano.tex     % Section 6
\input conserve.tex  % Section 7
\input related.tex   % related work, conclusion, acknowledgments

% Switch to the following when pulling from master.bib
%\bibliography{../../../Dropbox/Writing/references/master} % From git repo
%\bibliography{../references/master} % From Dropbox repo
% Switch to the following when using the locally extracted bib file
%\bibliography{master-extract}
\vspace{-1.5ex}
\input bib.tex

\end{document}

%% file: macros.tex
\usepackage{amsmath}
\usepackage{amstext}
\usepackage{framed}
\usepackage{amssymb}
\usepackage{array}
\usepackage{cite}
\usepackage{cmll}
\usepackage{graphicx} % for \rotatebox
\usepackage{hyperref}
\usepackage{listings} % for listing code
\usepackage{proof}
\usepackage{tikz}
\usepackage{wrapfig} % For the one figure-in-paragraph
\usepackage{xspace}

                     {\endgroup\end{quote}}

\newcommand{\blue}{\color{blue}}

\newcommand{\ie}{{\em i.e.}}
\newcommand{\eg}{{\em e.g.}}

\newcommand{\x}{\vec{x}}

\renewcommand{\t}{\vec{t}}
\newcommand{\B}{\overline{B}}

\newcommand{\tup}[1]{\langle#1\rangle}
\newcommand{\ra}{\rightarrow}
\newcommand{\lltens}{\mathrel{\otimes}}
\newcommand{\llone}{1}
\newcommand{\llzero}{0}
\newcommand{\llplus}{\mathrel{\oplus}}
\newcommand{\llpar}{\mathrel{\invamp}}

\newcommand{\llwith}{\mathrel{\&}}

\newcommand{\llimp}{\multimap}

\newcommand{\bang}{\mathop{!}} % DM Improves readability
\newcommand{\quest}{\mathop{?}} % DM Improves readability
\newcommand{\iimp}{\supset}
\newcommand{\TT}{\hbox{\textsl{tt}}}
\newcommand{\FF}{\hbox{\textsl{ff}}}
\newcommand{\limp}{\multimap}
\newcommand{\dual}[1]{{\color{blue}\overline{\color{black}#1}}} % Dual

\newcommand{\muMALL}    {$\mu\proofsystem{MALL}$\xspace}

\newcommand{\proofsystem}[1]{\mbox{\rm #1}\xspace}
\newcommand{\LK}        {\proofsystem{LK}}
\newcommand{\LKF}       {\proofsystem{LKF}}

\newcommand{\MALLF}     {\proofsystem{MALLF}}
\newcommand{\prefix}    {\bar{\bar{\mu}}}
\newcommand{\mumall}    {$\prefix\proofsystem{MALL}$\xspace}
\newcommand{\mlk}       {$\prefix\hbox{LK}_p$\xspace}
\newcommand{\mlkp}      {$\prefix\hbox{LK}_p^{+}$\xspace}

\newcommand{\mlkNeg}    {$\prefix\hbox{LK}_p(\Neg_1)$\xspace}
\newcommand{\mlkNegs}   {$\prefix\hbox{LK}_p^*(\Neg_1)$\xspace}

\newcommand{\mumallx}   {$\prefix\hbox{MALL}(\Bang,\Quest)$\xspace}

\newcommand{\Pos}{{\bf P}}
\newcommand{\Neg}{{\bf N}}

\newcommand{\oneside}[1]{\vdash #1}
\newcommand{\twoside}[2]{#1 \vdash #2}

\newcommand{\init}{\hbox{\textsl{init}}\xspace}
\newcommand{\cut}{\hbox{\textsl{cut}}\xspace}

\newcommand{\unfold}{\hbox{\textsl{unfold}}\xspace}

\newcommand{\induction}{\hbox{\sl induction\/}}
\newcommand{\coinduction}{\hbox{\sl coinduction\/}}

\newcommand{\Bang }{\mathord{\bang}}
\newcommand{\Quest}{\mathord{\quest}}

\newcommand{\emp}{\cdot}
\newcommand{\state}[3]{\tup{#1\,;~ #2\,;~#3}}
\newcommand{\cbar}[2]{#1 {\color{red}\,\mathrel{|}\,#2}}
\newcommand{\trans}[2]{#1\Rightarrow #2}
\newcommand{\compose}{\mathbin{\circ}}

\newcommand{\s}[1]{(s~#1)}
\newcommand{\Zero}{{\bf 0}}
\newcommand{\One }{{\bf 1}}
\newcommand{\Two }{{\bf 2}}
\newcommand{\nat }{\hbox{\sl nat}}

\newcommand{\plus}{\hbox{\sl plus\/}}
\newcommand{\mult}{\hbox{\sl mult\/}}
\newcommand{\ack}{\hbox{\sl ack\/}}

\newcommand{\Cnu}[1]{C\nu_{#1}}

\newcommand{\pnproof}{P/N-proof\xspace}
\newcommand{\pnproofs}{P/N-proofs\xspace}

%% file: intro.tex
\section{Introduction}
\label{sec:intro}

First-order logic formulas are built from propositional connectives,
first-order quantifiers, first-order terms, and the class of
non-logical constants called \emph{predicates} that denote relations
between terms. Moving from first-order logic to first-order
arithmetic, one introduces induction principles and banishes the
predicate constants by formally defining relations between terms using
those inductive principles. When moving from classical logic to
arithmetic in this fashion, one arrives at a presentation of Peano
Arithmetic. In this paper, we continue the project of studying
arithmetic based instead on linear logic, which was initiated
in~\cite{baelde07lpar,baelde08phd,baelde12tocl}, and where this
linearized version of arithmetic was called \muMALL. Since that earlier work,
many researchers (\eg \cite{das22fscd,horne23places}) have adopted
this name to refer to the propositional fragment of this logic, \ie,
the fragment without first-order terms, quantification, and
equality. To emphasize our focus on first-order structures, we will
use the name \mumall (and \mlk) in the rest of this paper.

Linear logic has played various roles in computational logic. Many
applications rely on the ability of linear logic to capture the
multiset rewriting paradigm~\cite{banatre93cacm}, which, in turn, can
encode Petri nets~\cite{gehlot90lics}, process
calculi~\cite{kobayashi95fac,miller92welp}, and stateful
computations~\cite{hodas94ic,miller96tcs}. Our use of linear logic
here will have none of that flavor. While the sequent calculus we use
to present \mumall in Section~\ref{sec:proofs} is based on multisets
of formulas, we shall not model computation as some rewriting of
multisets of atomic-formulas-as-tokens. In contrast, when we use
linear logic connectives within arithmetic, we capture computation and
deduction via familiar means that rely on relations between numerical
expressions. We propose linearized arithmetic not to build a
non-standard arithmetic but to better understand computation and
reasoning in arithmetic.

Since we are interested in using \mumall to study \emph{arithmetic},
we use first-order structures to encode natural numbers and fixed
points to encode relations among numbers. This focus contrasts the
uses of the propositional subset of \mumall as a typing system (see,
for example, \cite{ehrhard21lics}). We shall limit ourselves to using
invariants to reason inductively about fixed points instead of
employing other methods, such as infinitary proof systems (e.g.,
\cite{brotherston11jlc}) and cyclic proof systems (e.g.,
\cite{simpson17fossacs,das19lmcs}).

We begin our analysis of arithmetic by demonstrating that functions
defined relationally in \mumall can be directly computed from their
specifications using unification and backtracking search
(Section~\ref{sec:proof search}).  We then introduce a new proof
system, \mlkp, which extends \mumall with the weakening and
contraction rules.  While the addition of these rules provides a
natural foundation for classical logic, the precise nature of \mlkp is
not well understood.  In particular, we do not yet know if \mlk, the
cut-free version of \mlkp, is equivalent to \mlkp or if it admits a
complete focusing proof system.  In this paper, we establish the
consistency of \mlkp, demonstrate its capacity to encode Peano
Arithmetic (Section~\ref{sec:peano}), and prove specific
conservativity results of \mlk over \mumall (Section~\ref{sec:la}).

% LocalWords:  lin eaire une malvenue qui sugg un syst eme alors les
% LocalWords:  qu'il s'agit permettant abandonner tous

%% file: terms.tex
\section{Terms and formulas}
\label{sec:terms}

We use Church's approach \cite{church40} to define terms, formulas,
and abstractions over these by making them all simply typed
$\lambda$-terms.  The primitive type $o$ denotes formulas (of linear
and classical logics).  In this paper, we assume that
there is a second primitive type $\iota$ and that the (ambient)
signature contains the constructors $z\colon\iota$ (zero)
and $s\colon\iota\ra\iota$ (successor).  We abbreviate the terms $z$,
$\s{z}$, $\s{\s{z}}$, $\s{\s{\s{z}}}$, etc by \Zero, \One, \Two, {\bf 3}, etc.

\subsection{Logical connectives involving type $\iota$}
\label{ssec:iota}

We first present the logical connectives that relate to first-order
structures.  The two quantifiers $\forall$ and $\exists$ are both
given the type $(\iota\ra o)\ra o$: the terms $\forall(\lambda x.B)$
and $\exists(\lambda x.B)$ of type $o$ are abbreviated as $\forall
x.B$ and $\exists x.B$, respectively.  Equality $=$ and non-equality
$\neq $ are both of the type $\iota\ra\iota\ra o$.  For $n\ge 0$, the
least fixed point operator of arity $n$ is written as $\mu_n$ and the
greatest fixed point operator of arity $n$ is written as $\nu_n$, and
they both have the type $(A\ra A)\ra A$ where $A$ is the type
$\iota\ra\cdots\ra \iota\ra o$ in which there are $n$ occurrences of
$\iota$.  We seldom write explicitly the arity of fixed points 
as it can usually be determined from context when its value is
important.  The pairs of connectives $\langle\forall,\exists\rangle$,
$\langle\mu,\nu\rangle$, and $\langle=,\neq\rangle$ are De Morgan
duals.

Our formalizations of arithmetic do not contain predicate constants:
we do not admit any non-logical symbols of type
$\iota\ra\cdots\ra\iota\ra o$.  Consequently, there are no atomic
formulas in the traditional sense, \ie, formulas headed by non-logical
symbols. Equality, non-equality, and fixed-point operators are treated
as logical connectives, as they will be given introduction rules in
the sequent calculus proof systems we will soon introduce.

We shall use the usual rules for $\lambda$-conversion, namely,
$\alpha$, $\beta$, and $\eta$ conversion~\cite{church40}, as equality
on both terms and formulas.  In general, we assume that terms and
formulas are in $\beta$-normal form. 

\subsection{Propositional connectives of linear logic}
\label{ssec:prop}

The eight linear logic connectives for MALL are the following.
\begin{center}
\begin{tabular}{ c|c|c|c|c } 
                & ~conjunction~ &~true~ &~disjunction~ &~false \strut\\ 
 \hline
 multiplicative~ & $\lltens$   & $\llone$ & $\llpar$  & $\bot$ \strut\\
 additive        & $\llwith$   & $\top$   & $\llplus$ & $\llzero$ \strut
\end{tabular}
\end{center}
The four binary connectives have type $o\ra o\ra o$, and the four
units have type $o$.  (The use of 0 and 1 as logical connectives is
unfortunate for a paper about arithmetic. As we mentioned above,
numerals are written in boldface.)  Formulas involving the set of
logical connectives in Section~\ref{ssec:iota} and these propositional
connectives are called \mumall formulas, a logic that was first
proposed in~\cite{baelde07lpar}. Many of the proof-theoretic
properties of \mumall will be summarized in Section~\ref{sec:proofs}.

We do not treat negation as a logical connective: when $B$ is
a formula, we write $\overline{B}$ to denote the formula resulting
from taking the De Morgan dual of $B$.  We occasionally use the linear
implication $B\limp C$ as an abbreviation for $\overline{B}\llpar C$.
We also use this overline notation when $B$ is the body of a fixed
point expression, \ie, when $B$ has the form $\lambda p\lambda\x. C$
where $C$ is a formula, $p$ is a first-order predicate variable, and
$\x$ is a list of first-order variables, then $\overline{B}$ is
$\lambda p\lambda\x. \overline{C}$~\cite[Definition
  2.1]{baelde12tocl}.  For example, if $B$ is \([\lambda p\lambda x. x
  = z \llplus\exists y. x = \s{\s{y}}\lltens p~y]\) then
$\overline{B}$ is \([\lambda p\lambda x. x \neq z \llwith\forall y. x
  \neq \s{\s{y}}\llpar p~y]\).

\subsection{Polarized and unpolarized formulas}
\label{ssec:polarized}

The connectives of linear logic are given a \emph{polarity} as
follows.  The \emph{negative} connectives are $\llpar$, $\bot$,
$\llwith$, $\top$, $\forall$, $\neq$, and $\nu$ while their De Morgan
duals---namely, $\lltens$, $\llone$, $\llplus$, $\llzero$, $\exists$,
$=$, and $\mu$---are positive.  A \mumall formula is positive or
negative depending only on the polarity of its topmost connective.
The polarity flips between $B$ and $\overline{B}$.  
We shall also call \mumall formulas \emph{polarized formulas}.

\emph{Unpolarized formulas} are built using the four usual classical
propositional logic connectives $\land$, $\TT$, $\lor$, $\FF$ plus 
$=$, $\not=$, $\forall$, $\exists$, $\mu$, and $\nu$.  Thus,
the six connectives with $i$ in their typing can appear in
polarized and unpolarized formulas.  Unpolarized formulas are
also called \emph{classical logic formulas}.  Note that unpolarized
formulas do not contain negations.  We shall extend the notation
$\overline{B}$ to unpolarized formulas $B$ in the same sense
as used with polarized formulas.  For convenience, we will
occasionally allow implications in unpolarized formulas: in those
cases, we treat $P\supset Q$ as $\overline{P}\lor Q$.

A polarized formula $\hat Q$ is a \emph{polarized version} of the
unpolarized formula $Q$ if every occurrence of $\llwith$ and $\lltens$
in $\hat Q$ is replaced by $\wedge$ in $Q$, every occurrence of
$\llpar$ and $\llplus$ in $\hat Q$ is replaced by $\vee$ in $Q$, every
occurrence of $\llone$ and $\top$ in $\hat Q$ is replaced by $\TT$ in
$Q$, and every occurrence of $\llzero$ and $\bot$ in $\hat Q$ is
replaced by $\FF$ in $Q$.  Notice that if $Q$ has $n$ occurrences of
propositional connectives, then there are $2^n$ formulas $\hat Q$ that
are polarized versions of $Q$.

Fixed point expressions, such as $((\mu\lambda P\lambda x
(B~P~x))~t)$, introduce variables of predicate type (here, $P$) into
their scope.  In the case of the $\mu$ fixed point, any formula built
using that predicate variable as its topmost symbol will be
considered positively polarized.  Dually, if the $\nu$ operator is
used instead, any formula built using the predicate variable it
introduces is considered negatively polarized.  For example,
expression $(p~y)$ in \([\mu\lambda p\lambda x. x = z \llplus\exists
  y. x = \s{\s{y}}\lltens p~y]\) is polarized positively while that
same expression in \([\nu\lambda p\lambda x. x \neq z \llwith\forall
  y. x \neq \s{\s{y}}\llpar p~y]\) is polarized negatively.

\subsection{The polarization hierarchy}
\label{ssec:hierarchy}

A formula is \emph{purely positive} (resp., \emph{purely negative}) if
every logical connective it contains is positive (resp., negative).
Taking inspiration from the familiar notion of the arithmetical
hierarchy, we define the following collections of formulas. The
formulas in $\Pos_1$ are the purely positive formulas, and the
formulas in $\Neg_1$ are the purely negative formulas.  More
generally, for $n\ge 1$, the $\Neg_{n+1}$-formulas are those negative
formulas for which every occurrence of a positive subformula is a
$\Pos_n$-formula.  Similarly, the $\Pos_{n+1}$-formulas are those
positive formulas for which every occurrence of a negative subformula
is an $\Neg_n$-formula.  A formula in $\Pos_n$ or in $\Neg_n$ has at
most $n-1$ alternations of polarity.  Clearly, the dual of a
$\Pos_n$-formula is an $\Neg_n$-formula, and vice versa.  We shall also
extend these classifications of formulas to abstractions over terms:
thus, we say that the term $\lambda x. B$ of type $i\ra o$ is in
$\Pos_n$ if $B$ is a $\Pos_n$-formula.

Note that for all $n\ge 1$, if $B$ is an unpolarized $\Pi_n^0$-formula
(in the usual arithmetic hierarchy) then there is a polarized version of
$B$ that is $\Neg_n$.  Similarly, if $B$ is an unpolarized
$\Sigma_n^0$-formula then there is a polarized version of $B$ that is
$\Pos_n$.

% LocalWords:  overline

%% file: polarized.tex
\section{Linear and classical proof systems for polarized formulas}
\label{sec:proofs}

\subsection{The \mumall and \mlkp proof systems}
\label{sec:mumall}

The \mumall proof system \cite{baelde07lpar,baelde12tocl} for
polarized formulas is the one-sided sequent calculus proof system
given in Figure~\ref{fig:mumall}.  The variable $y$ in the
$\forall$-introduction rule is an \emph{eigenvariable}: it is
restricted from appearing free in any formula in the conclusion of that
rule.  In the $\neq$-introduction rule, if the terms $t$ and $t'$ are
not unifiable, the premise is empty and the conclusion follows
immediately.

The choice of using Church's $\lambda$-notation provides an elegant
treatment of higher-order substitutions (needed for handling induction
invariants) and provides a simple treatment of fixed point expressions
and the binding mechanisms used there.  In particular, we shall assume
that formulas in sequents are always treated modulo
$\alpha\beta\eta$-conversion.  We usually display formulas in
$\eta$-long, $\beta$-normal form when presenting sequents.  Note that
formula expressions such as $B~S~\t$ (see Figure~\ref{fig:mumall}) are
parsed as $(\cdots((B~S) t_1) \cdots t_n)$ if $\t$ is the list of
terms $t_1, \ldots, t_n$.

\begin{figure}[t]
\[ %  \arraycolsep=1.4pt\def\arraystretch{1.0}
\begin{array}{c@{\strut\qquad}c@{\strut\qquad}c@{\strut\qquad}c}
  \infer[\llpar]{\oneside{\Gamma, B\llpar C}}{\oneside{\Gamma, B, C}}
  &
  \infer[\bot]{\oneside{\Gamma, \bot}}{\oneside{\Gamma}}
  &
  \infer[\lltens]{\oneside{\Gamma,\Delta, B\lltens C}}
                 {\oneside{\Gamma, B} & \oneside{\Delta, C}}
  &
  \infer[\llone]{\oneside{\llone}}{\strut}
\\
  \infer[\llwith]{\oneside{\Gamma, B \llwith C}}
                 {\oneside{\Gamma, B} & \oneside{\Gamma, C}}
  &
  \infer[\top]{\oneside{\Gamma, \top}}{\strut}
  &
  \infer[\llplus]{\oneside{\Gamma, B_0\llplus B_1}}{\oneside{\Gamma, B_i}}
  \\
  \infer[\neq]{\oneside{\Gamma, t\neq t'}}
              {\{~\oneside{\Gamma\theta : \theta= mgu(t,t')~\}}}
  &
  \infer[=]{\oneside{t=t}}{\strut}
  &
  \infer[\forall]{\oneside{\Gamma, \forall x. B x}}
                 {\oneside{\Gamma, B y}}
  &
  \infer[\exists]{\oneside{\Gamma, \exists x. B x}}
                 {\oneside{\Gamma, Bt}}
\end{array}
\]
\[\begin{array}{c@{\strut\qquad}c@{\strut\qquad}c@{\strut\qquad}c}
  \infer[\nu]{\oneside{\Gamma, \nu B\t}}
             {\oneside{\Gamma, S\t} & \oneside{BS\x, \overline{(S\x)}}}
  &
  \infer[\mu]{\oneside{\Gamma, \mu B\t}}{\oneside{\Gamma, B(\mu B)\t}}
  &
  \infer[\mu\nu]{\oneside{\mu B\t, \nu \B\t}}{\strut}
\end{array}
\]
\caption{The inference rules for the \mumall proof system}
\label{fig:mumall}
\[
  \infer[\unfold]{\oneside{\Gamma, \nu B\t}}
                   {\oneside{\Gamma, B(\nu B)\t}}
  \qquad
  \infer[\init]{\oneside{B,\overline{B}}}{}
  \qquad
  \infer[\cut]{\oneside{\Gamma,\Delta}}
             {\oneside{\Gamma,B}\quad\oneside{\Delta,\overline{B}}}
\]
\caption{Three rules admissible in \mumall}
\label{fig:three}
\[
  \infer[C]{\oneside{\Gamma, B}}{\oneside{\Gamma, B, B}}
  \qquad
  \infer[W]{\oneside{\Gamma, B}}{\oneside{\Gamma}}
\]
\caption{Two structural rules}
\label{fig:structural}
\end{figure}

If we were working in a two-sided calculus, the $\nu$-rule in
Figure~\ref{fig:mumall} would split into the two rules
\[
\infer[\coinduction\qquad\hbox{and}\qquad]
      {\twoside{\Gamma}{\nu B\t,\Delta}}{
       \twoside{\Gamma}{\Delta,S\t}\quad\twoside{S\x}{B S \x}}
\infer[\induction.]{\twoside{\Gamma,\mu B\t}{\Delta}}{
         \twoside{\Gamma,S\t}{\Delta}\quad\twoside{B S\x}{S \x}}
\]
That is, the rule for $\nu$ yields both coinduction and induction.
In general, we shall speak of the higher-order substitution term $S$
used in both of these rules as the \emph{coinvariant} of that rule.

We make the following observations about this proof system.
\begin{enumerate}
\item The $\mu$ rule allows the $\mu$ fixed point to be unfolded.
  This rule captures, in part, the identification of $\mu B$ with
  $B(\mu B)$; that is, $\mu B$ is a fixed point of $B$.  This
  inference rule allows one occurrence of $B$ in $(\mu B)$ to be
  expanded to two occurrences of $B$ in $B(\mu B)$.  In this way,
  unbounded behavior can appear in \mumall where it does not occur in
  MALL.

\item The proof rules for equality guarantee that function symbols are
  all treated injectively; thus, function symbols will act only as
  term constructors.  In this paper, the only function symbols we
  employ are for zero and successor: of course, a theory of arithmetic
  should treat these symbols injectively.

\item The admissibility of the three rules in
  Figure~\ref{fig:three} for \mumall is proved in
  \cite{baelde12tocl}.  The general form of the initial rule is
  admissible, although the proof system only dictates a limited form
  of that rule via the $\mu\nu$ rule.  The $\unfold$ rule in
  Figure~\ref{fig:three}, which simply unfolds $\nu$-expression,
  is admissible in \mumall by using the $\nu$-rule with the coinvariant
  $S=B(\nu B)$.

\item While the weakening and contraction rules are not generally
  admissible in \mumall, they are both admissible for
  $\Neg_1$-formulas, a fact that plays an important role in
  Section~\ref{sec:la}.
\end{enumerate}

We could add the inference rules for equality, non-equality, and least
and greatest fixed points to Gentzen's LK proof system for first-order
classical logic \cite{gentzen35}.  We take a different approach,
however, in that, we will only consider proof systems for classical
logic using polarized versions of classical formulas.  In particular,
the \mlkp proof system is the result of adding to the \mumall proof
system the rules for contraction $C$ and weakening $W$ from
Figure~\ref{fig:structural} as well as the cut rule.

\subsection{Examples}
\label{sec:examples}

The formula $\forall x\forall y [x=y\lor x\neq  y]$ can be polarized
as either
\[
  \forall x\forall y [x=y\llpar  x\neq  y]\hbox{\quad or \quad }
  \forall x\forall y [x=y\llplus x\neq  y].
\]  
These polarized formulas belong to $\Neg_2$ and $\Neg_3$, respectively.
Only the first of these is provable in
\mumall, although both formulas are provable in \mlk.

Note that it is clear that if there exists a \mumall proof of a
$\Pos_1$-formula, then that proof does not contain the $\nu$ rule,
i.e., it does not contain the rules involving coinvariants.  Finally,
given that first-order Horn clauses can interpret Turing
machines~\cite{tarnlund77}, and given that Horn clauses can easily be
encoded using $\Pos_1$-formulas, it is undecidable whether or not a
$\Pos_1$ expression has a \mumall proof.  Similarly, $\Pos_1$
formulas can be used to specify any general recursive function.
Obviously, the provability of $\Neg_1$-formulas is also undecidable.

The following are proofs of 
two axioms of Peano Arithmetic (see also Section~\ref{sec:peano}).

\[
  \infer[\forall]{\oneside{\forall x.~(s\, x) \neq \Zero}}{
  \infer[\neq]{\oneside{(s\, x) \neq \Zero}}{}}
  \qquad
  \infer[\forall\times 2]
        {\oneside{\forall x \forall y.~(s\, x = s\, y)\supset(x = y)}}{
  \infer[\llpar]{\oneside{(s\,x \neq s\,y)\llpar(x = y)}}{
  \infer[\neq]{\oneside{s\,x \neq s\,y,x = y}}{
  \infer[=]{\oneside{x = x}}{
}}}}
\]

The unary relation for denoting the set of natural numbers and the
ternary relations for addition and multiplication can be 
axiomatized using Horn clauses as follows.
\begin{align*}
  &\nat~\Zero\\
\forall x(\nat~x \supset\;&\nat~\s{x})\\[6pt]
\forall x (&\plus~\Zero~x~x)\\
\forall x\forall y\forall u (\plus~x~y~u\supset\; &\plus~\s{x}~y~\s{u})\\[6pt]
\forall x (&\mult~\Zero~x~\Zero)\\
\forall x \forall y\forall u\forall w (\mult~x~y~u\land \plus~y~u~w\supset\; &\mult~\s{x}~y~w)
\end{align*}
These Horn clauses can be mechanically transformed into the following
least fixed point definitions of these relations.
\[
  \nat = \mu\lambda N\lambda x (x=\Zero \oplus \exists x'(x=\s{x'}\lltens N~x'))
\]
\[
  \plus = \mu\lambda P\lambda x\lambda y\lambda u
       ((x=\Zero\lltens y=u)\llplus
        \exists x'\exists u'\exists w(x=\s{x'}\lltens u=\s{u'}\lltens P~x'~y~u'))
\]
\[
  \mult = \mu\lambda M\lambda x\lambda y\lambda w
          \big((x=\Zero\lltens u=\Zero)\llplus 
               \exists x'\exists u'\exists w(x=\s{x'}\lltens plus~y~u'~w
                                            \lltens M~x'~y~w)
\big)
\]
Both of these fixed point expressions are $\Pos_1$.

The following derivation verifies that 4 is a sum of 2 and 2.
\[
  \infer[\exists,\lltens]{\oneside{\exists p. \plus~\Two~\Two~p\lltens\nat~p}}{
  \infer[\mu]{\oneside{\plus~\Two~\Two~{\bf 4}}}{
  \infer[\llplus]{\oneside{(\Two=\Zero\lltens \Two={\bf 4})\llplus
        \exists n'\exists p'(\Two=\s{n'}\lltens {\bf 4}=\s{p'}\lltens
        P~n'~\Two~p')}}{
  \infer=[\exists\times 2]{\oneside{\exists n'\exists p'(\Two=\s{n'}\lltens 
        {\bf 4}=\s{p'}\lltens \plus~n'~\Two~p')}}{
  \infer=[\lltens\times 2]{\oneside{\Two=\s{\One}\lltens {\bf 4}=\s{{\bf 3}}\lltens
         \plus~\One~\Two~{\bf 3}}}{
         \infer[=]{\oneside{\Two=\s{\One}}}{} \quad
         \infer[=]{\oneside{{\bf 4}=\s{{\bf 3}}}}{} \quad
         \oneside{\plus~\One~\Two~{\bf 3}}
  }}}} &
  \oneside{\nat~{\bf 4}}}
\]
To complete this proof, we must construct the (obvious) proof of
$\oneside{\nat~{\bf 4}}$ and a similar subproof verifying
that $1+2=3$.  Note that in the bottom-up construction of this proof,
the witness used to instantiate the final $\exists p$ is, in fact, the
sum of 2 and 2.  Thus, this proof construction does not compute this
sum's value but simply checks that 4 is the correct value.

In contrast to the above example, the following proof of 
$\forall u (\plus~\Two~\Two~u \supset \nat~u)$
can be seen as a \emph{computation} of the value of 2 plus 2.  The
proof of this sequent begins as follows.
\[
  \infer[\forall,\llpar]
        {\oneside{\forall u (\overline{\plus~\Two~\Two~u}\llpar\nat~u)}}{
  \infer[\kern -2.5pt\unfold,\with]
        {\oneside{\overline{\plus~\Two~\Two~u},\nat~u}}{
         \infer[\kern -4pt\llpar,\neq]
               {\oneside{\Two\neq\Zero\llpar \Two\neq u,\nat~u}}{}~~
         \infer[\forall,\llpar]
               {\oneside{\forall x'\forall u'\forall w(\Two\neq\s{x'}\llpar
                                                       u\neq\s{u'}\llpar 
                \overline{\plus~x'~\Two~u'}),\nat~u}}{
          \infer[\neq\times 2]
                {\oneside{\Two\neq\s{x'}, u\neq\s{u'}, 
                   \overline{\plus~x'~\Two~u'},\nat~u}}{
                 \oneside{\overline{\plus~\One~\Two~u'},\nat~\s{u'}}}}}}
\]
Similarly, the open premise above has a partial proof which reduces
its provability to the provability of the sequent
$\oneside{\overline{\plus~\Zero~\Two~u'},\nat~\s{\s{u'}}}$.  This
final sequent is similarly reduced to $\oneside{\Zero\neq{\bf
    0},\Two\neq u',\nat~\s{\s{u'}}}$, which is itself reduced to
$\oneside{\nat~{\bf 4}}$, which has a trivial proof.  Note that the
bottom-up construction of this proof involves the systematic
computation of the value of 2 plus 2.

The previous two proofs involved with the judgment $2+2=4$ illustrates
two different ways to determine $2+2$: the first involves a
``guess-and-check'' approach, while the second involves a direct
computation.  We will return to these two approaches in
Section~\ref{sec:proof search}.

Unpolarized formulas that state the \emph{totality} and
\emph{determinacy} of the function encoded by a binary relation $\phi$
can be written as
\begin{align*}
  [\forall x .&\nat~x\supset\exists y. \nat~y \land\phi(x,y)]\;\land \\
  [\forall x .&\nat~x\supset\forall y_1.\nat~y_1\supset \forall y_2.\nat~y_2\supset 
      \phi(x,y_1) \supset \phi(x,y_2) \supset y_1=y_2].
\end{align*}
If this formula is polarized so that the two implications are encoded
using $\llpar$, the conjunction is replaced by $\with$, and the
expression $\phi$ is $\Pos_1$, then this formula is an $\Neg_2$
formula.

Given the definition of addition on natural numbers above, the
following totality and determinacy formulas
\begin{align*}
  [\forall x_1\forall x_2.&~\nat~x_1\supset\nat~x_2\supset
              \exists y.(\plus(x_1,x_2,y)\land\nat~y)] \\
  [\forall x_1\forall x_2.&~\nat~x_1\supset\nat~x_2\supset\forall y_1\forall y_2.~
     \plus(x_1,x_2,y_1)\supset \plus(x_1,x_2,y_2) \supset y_1=y_2]
\end{align*}
can be proved in \mumall where $\supset$ is polarized using $\llpar$
and the one occurrence of conjunction above is polarized using
$\with$.  These proofs require both induction and the $\mu\nu$ rule.

The direct connection between proof search in \mumall and the model
checking problems of reachability and bisimilarity (and their
negations) has been demonstrated in \cite{heath19jar}.  In particular,
reachability problems were encoded as $\Pos_1$-formulas, while
non-reachability problems were encoded as $\Neg_1$-formulas.  The
paper \cite{heath19jar} also showed that the specification of
simulation and bisimulation can be encoded as $\Neg_2$- formulas.
Another common form of $\Pos_1$-formulas arises when applying the Clark
completion~\cite{clark78} to Horn clause specifications.

\subsection{Some known results concerning \mumall}
\label{ssec:results}

While \mumall does not contain the contraction rule, it is still
possible for the number of occurrences of logical connectives to grow
in sequents when searching for a proof.  In particular, the unfolding
rule (when read from conclusion to premise) can make a sequent
containing $(\mu B\t)$ into a sequent containing $(B(\mu B)\t)$: here,
the abstracted formula $B$ is repeated.  Surprisingly, however, 
the subset of \mumall that does not contain occurrences of fixed
points is still undecidable.  In particular, consider the
following two sets of inductively defined classes of \mumall formulas.
\[
\begin{array}{cl}
  \Phi~::=&
          \Phi \with \Phi       \; | \; 
          \exists x. \Phi       \; | \; 
          \forall x. \Phi       \; | \; 
          \Psi\\
  \Psi~::=& 
          t_1 = t_1' \limp\cdots\limp t_n = t_n' \limp t_0 = t_0'
          \quad(n\ge 0)
\end{array}
\]
If we also assume that there are exactly three constructors, one each
of type $\iota\ra\iota$, $\iota\ra\iota\ra\iota$, and
$\iota\ra\iota\ra\iota\ra\iota$, then it is undecidable whether or not
a given formula $\Phi$ is provable in \mumall~\cite{miller22amai}.

The two main proof-theoretic results concerning \mumall 
are the admissibility of the cut rule (in Figure~\ref{fig:three})
and the completeness of a focusing proof system~\cite{baelde12tocl}.

\subsection{Definable exponentials}
\label{ssect:exponentials}

As Baelde showed in \cite{baelde12tocl}, the following definitions 
\[
  \Quest P = \mu(\lambda p. \bot\oplus (p\llpar p)\oplus P)
  \qquad
  \Bang P = \dual{\Quest{(\dual{P})}}
\]
approximate the exponentials of linear logic in the sense that the
following four rules---dereliction, contraction, weakening, and
promotion---are admissible in \mumall. 
\[
  \infer[W]{\oneside{\Quest B, \Gamma}}{\oneside{\Gamma}}
  \qquad
  \infer[C]{\oneside{\Quest B, \Gamma}}{\oneside{\Quest B,\Quest B,\Gamma}}
  \qquad
  \infer[D]{\oneside{\Quest B, \Gamma}}{\oneside{B,\Gamma}}
  \qquad
  \infer[P]{\oneside{\Bang B, \Quest\Gamma}}{\oneside{B,\Quest\Gamma}}
\]
In particular, we use \mumallx to denote the extension of \mumall with
the two exponentials $\Bang$ and $\Quest$ and the above four proof
rules.  Thus, every \mumallx-provable sequent can be mapped to a
\mumall-provable sequent by simply replacing the exponentials for
their corresponding fixed point definition.

% LocalWords:  Bt unary

%% file: compute.tex
\section{Using proof search to compute functions}
\label{sec:proof search}

We say that a binary relation $\phi$ encodes a function $f$ if
$\phi(x,y)$ holds if and only if $f(x)=y$.  Of course, this
correspondence is only well-defined if we know that the \emph{totality}
and \emph{determinacy} properties hold for $\phi$.
For example, let $\plus$ be the definition of addition on natural
numbers given in Section~\ref{sec:proofs}.  The following polarized
formulas encoding totality and determinacy are $\Neg_2$-formulas.
\begin{align*}
  [\forall x_1\forall x_2.&~\nat~x_1\limp\nat~x_2\limp
              \exists y.\nat~y\lltens\plus~x_1~x_2~y] \\
  [\forall x_1\forall x_2.&~\nat~x_1\limp\nat~x_2\limp
   \forall y_1\forall y_2.~ \plus~x_1~x_2~y_1\limp 
                      \plus~x_1~x_2~y_2 \limp y_1=y_2]
\end{align*}
These formulas can be proved in \mumall. 

% Different computational approaches

One approach to computing the function that adds two natural numbers
is to follow the Curry-Howard approach of relating proof theory to
computation~\cite{howard80}.   First, extract from a natural
deduction proof of the totality formula above a typed $\lambda$-term.
Second, apply that $\lambda$-term to the $\lambda$-terms representing
the two proofs of, say, $\nat~n$ and $\nat~m$.  Third, use a
nondeterministic rewriting process that iteratively selects
$\beta$-redexes for reduction.  In most typed $\lambda$-calculus
systems, all such sequences of rewritings will end in the same normal
form, although some sequences of rewrites might be very long, and
others can be very short.  The resulting normal $\lambda$-term should
encode the proof of $\nat~p$, where $p$ is the sum of $n$ and $m$.
In this section, we will present an alternative mechanism for
computing functions from their relational specification that relies on
using proof search mechanisms instead of this proof-normalization
mechanism.

% Singleton observation

The totality and determinacy properties of some binary relation $\phi$
can be expressed equivalently as, for any natural number $n$, the
expression $\lambda y.\phi(n,y)$ denotes a singleton set.  Of course,
the sole member of that singleton set is the value of the function it
encodes.  If our logic contained a choice operator, such as Church's
\emph{definite description} operator $\iota$~\cite{church40}, 
this function can be represented as $\lambda x.\iota
y. \phi(x,y)$.  The search for proofs can, however, be used to provide
a more computational approach to computing the function encoded by
$\phi$.  Assume that $P$ and $Q$ are predicates of arity one and that
$P$ denotes a singleton.  In this case, the (unpolarized) formulas
$\exists x[P x\land Q x]$ and $\forall x[P x\supset Q x]$ are
logically equivalent, although the proof search semantics of these
formulas are surprisingly different.  In particular, if we attempt to
prove $\exists x[P x\land Q x]$, then we must \emph{guess} a term $t$
and then \emph{check} that $t$ denotes the element of the singleton
(by proving $P(t)$).  In contrast, if we attempt to prove $\forall x[P
  x\supset Q x]$, we allocate an eigenvariable $y$ and attempt to
prove the sequent $\vdash P y\supset Q y$.  Such an attempt at
building a proof might \emph{compute} the value $t$ (especially if we
can restrict proofs of that implication not to involve the general
form of induction).  This difference was illustrated in
Section~\ref{sec:proofs} with the proof of $\oneside{\exists
  p. \plus~\Two~\Two~p\lltens\nat~p}$ (which guesses and checks that
the value of 2 plus 2 is 4) versus the proof of $\forall u
(\overline{\plus~\Two~\Two~u}\llpar\nat~u)$ (which incrementally
constructs the sum of 2 and 2).

Assume that $P$ is a $\Pos_1$ predicate expression of type
$i\ra o$ and that we have a \mumall proof of $\forall x[P x\supset
  \nat~x]$.  If this proof does not contain the induction rule, then
that proof can be seen as computing the sole member of $P$.  As the
following example shows, it is not the case that if there is a \mumall
proof of $\forall x[P x\supset \nat~x]$ then it has a proof in which
the only form of the induction rule is unfolding.
To illustrate this point, 
let $P$ be $\mu(\lambda R\lambda x. x = \Zero \oplus (R~\s{x}))$.
Clearly, $P$ denotes the singleton set containing zero.  There is also
a \mumall proof that $\forall x[P x\supset \nat~x]$, but there is no
(cut-free) proof of this theorem that uses unfolding instead of the
more general induction rule: just using
unfoldings leads to an unbounded proof search attempt, which
follows the outline
\[
  \infer=[\unfold,\llwith,\neq.]
         {\oneside{\overline{P~y},\nat~y}}{\oneside{\nat~\Zero} &
  \infer=[\unfold,\llwith,\neq]
         {\oneside{\overline{P~\s{y}},\nat~y}}{
  \deduce{\oneside{\overline{P~\s{\s{y}}},\nat~y}}{\vdots}}
}
\]

Although proof search can contain potentially unbounded branches, we
can still use the proof search concepts of unification and
nondeterministic search to compute the value within a singleton.  We
now define a nondeterministic algorithm to do exactly that.  The
\emph{state} of this algorithm is a triple of the form
\[\state{x_1,\ldots,x_n}{B_1,\ldots,B_m}{t},\] where $t$ is a term,
$B_1,\ldots,B_m$ is a multiset of $\Pos_1$-formulas, and all
variables free in $t$ and in the formulas $B_1,\ldots,B_m$ are in the set of
variables $x_1,\ldots,x_n$.  A \emph{success state} is one of the form
$\state{\emp}{\emp}{t}$ (that is, when $n=m=0$): such a state is
said to have \emph{value} $t$.

Given the state $S=\state{\Sigma}{B_1,\ldots,B_m}{t}$ with
$m\ge1$, we can nondeterministically select one of the $B_i$
formulas. For the sake of simplicity, assume that we have selected
$B_1$.  We define the transition $\trans{S}{S'}$ of state $S$ to state
$S'$ by a case analysis of the top-level structure of $B_1$.

\begin{itemize}
\item If $B_1$ is $u=v$ and the terms $u$ and $v$ are unifiable with
  most general unifier $\theta$, then we transition to
  \(\state{\Sigma\theta}{B_2\theta,\ldots,B_m\theta}{t\theta}.\)
\item If $B_1$ is $B\lltens B'$ then we transition to
  $\state{\Sigma}{B, B', B_2,\ldots,B_m}{t}$.
\item If $B_1$ is $B\llplus B'$ then we transition to either 
  \(\state{\Sigma}{B,  B_2,\ldots,B_m}{t}\) or
  \(\state{\Sigma}{B', B_2,\ldots,B_m}{t}\).
\item If $B_1$ is $\mu B \t$ then we transition to
  $\state{\Sigma}{B (\mu B) \t, B_2,\ldots,B_m}{t}$.
\item If $B_1$ is $\exists y.~ B~y$ then we transition to
  $\state{\Sigma, y}{B~y, B_2,\ldots,B_m}{t}$ assuming that $y$
  is not in $\Sigma$. 
\end{itemize}

This nondeterministic algorithm is essentially applying
left-introduction rules (assuming a two-sided sequent calculus) in a
bottom-up fashion and, if there are two premises, selecting
(nonde\-ter\-min\-is\-ti\-cal\-ly) just one premise to follow.

\begin{lemma}\label{lemma:compute}
Assume that $P$ is a $\Pos_1$ expression of type $i\ra o$ and
that $\exists y. P y$ has a \mumall proof.  There is a
sequence of transitions from the initial state
$\state{y}{P~y}{y}$ to a success state with value $t$ 
such that $P~t$ has a \mumall proof.
\end{lemma}

\begin{proof}
An \emph{augmented state} is a structure of the form  
\(
  \state{\cbar{\Sigma}{\theta}}
        {\cbar{B_1}{\Xi_1},\ldots,\cbar{B_m}{\Xi_m}}{t},
\)
where
\begin{itemize}
  \item $\theta$ is a substitution with domain equal to $\Sigma$ and
    which has no free variables in its range, and 
  \item for all $i\in\{1,\ldots,m\}$, $\Xi_i$ is a (cut-free) \mumall
    proof of $\theta(B_i)$. 
\end{itemize}
Note that we are left with a regular state if we strike out the augmented items.
Given that we have a \mumall proof of $\exists
y. P y$, we must have a \mumall proof $\Xi_0$ of $P~t$ for some term
$t$.  Note that there is no occurrence of the induction rule in
$\Xi_0$.  We now set the initial augmented state to
$\state{\cbar{y}{[y\mapsto t]}}{\cbar{P y}{\Xi_0}}{y}$.  As we detail
now, the proof structures $\Xi_i$ provide oracles that steer this
nondeterministic algorithm to a success state with value $t$.  Given
the augmented state
\[
  \state{\cbar{\Sigma}{\theta}}{\cbar{B_1}{\Xi_1},\ldots,\cbar{B_m}{\Xi_m}}{s},
\]
we consider selecting the first pair $\cbar{B_1}{\Xi_1}$ and consider
the structure of $B_1$.
\begin{itemize}

\item If $B_1$ is $B'\lltens B''$ then the last inference rule of
  $\Xi_1$ is $\lltens$ with premises $\Xi'$ and $\Xi''$, and we make a
  transition to 
  \( \state{\cbar{\Sigma}{\theta}}
           {\cbar{B'}{\Xi'},\cbar{B''}{\Xi''},\ldots,\cbar{B_m}{\Xi_m}}
           {s}\).

\item If $B_1$ is $B'\llplus B''$ then the last inference rule of
  $\Xi_1$ is $\llplus$, and that rule selects either the first or the second
  disjunct.  In either case, let $\Xi'$ be the proof of its premise. 
  Depending on which of these disjuncts is selected, we make a
  transition to either 
$\state{\cbar{\Sigma}{\theta}}{\cbar{B'}{\Xi'},
    \cbar{B_2}{\Xi_2},\ldots,\cbar{B_m}{\Xi_m}}{s}$ or 
$\state{\cbar{\Sigma}{\theta}}
        {\cbar{B''}{\Xi'}, \cbar{B_2}{\Xi_2},\ldots,\cbar{B_m}{\Xi_m}}{s}$,
respectively.

\item If $B_1$ is $\mu B \t$ then the last inference rule of $\Xi_1$ is
  $\mu$.  Let $\Xi'$ be the proof of the premise of that inference
  rule.  We make a transition to
  \(\state{\cbar{\Sigma}{\theta}}
          {\cbar{B (\mu B) \t}{\Xi'}, \cbar{B_2}{\Xi_2},\ldots,\cbar{B_m}{\Xi_m}}
          {s}\).

\item If $B_1$ is $\exists y.~B~y$ then the last inference rule of
  $\Xi_1$ is $\exists$.  Let $r$ be the substitution term used to
  introduce this $\exists$ quantifier and let $\Xi'$ be the proof of
  the premise of that inference rule.  Then, we make a transition to
  \(\state{\cbar{\Sigma,w}{\theta\compose\varphi}}
          {\cbar{B~w}{\Xi'}, \cbar{B_2}{\Xi_2},\ldots,\cbar{B_m}{\Xi_m}}
          {s}\),
  where $w$ is a variable not in $\Sigma$ and $\varphi$ is the
  substitution $[w\mapsto r]$.  Here, we assume that the composition
  of substitutions satisfies the equation $(\theta\compose\varphi)(x)
  = \varphi(\theta(x))$.

\item If $B_1$ is $u=v$ and the terms $u$ and $v$ are unifiable with
  most general unifier $\varphi$, then we make a transition to
  \(\state{\cbar{\Sigma\varphi}{\rho}}
          {\cbar{\varphi(B_2)}{\Xi_2},\ldots,\cbar{\varphi(B_m)}{\Xi_m}}
          {(\varphi t)}\)
  where $\rho$ is the substitution such that $\theta =
  \varphi\compose\rho$.
\end{itemize}

We must show that the transition is made to an augmented state in each
of these cases.  This is easy to show in all but the last two rules
above.  In the case of the transition due to $\exists$, we know that
$\Xi'$ is a proof of $\theta(B~r)$, but that formula is simply
$\varphi(\theta(B~w))$ since $w$ is new and $r$ contains no variables
free in $\Sigma$.  In the case of the transition due to equality, we
know that $\Xi_1$ is a proof of the formula $\theta (u=v)$, which means
that $\theta u$ and $\theta v$ are the same terms and, hence, that $u$
and $v$ are unifiable and that $\theta$ is a unifier.  Let $\varphi$
be the most general unifier of $u$ and $v$.  Thus, there is a
substitution $\rho$ such that $\theta = \varphi\compose\rho$ and, for
$i\in\{2,\ldots,m\}$, $\Xi_i$ is a proof of
$(\varphi\compose\rho)(B_i)$.  Finally, termination of this algorithm
is ensured since the number of occurrences of inference rules in the
included proofs decreases at every step of the transition.  Since we
have shown that there is an augmented path that terminates, we have
that there exists a path of states to a success state with value $t$.
\end{proof}

This lemma ensures that our search algorithm can compute a member from
a non-empty set, given a \mumall proof that that set is non-empty.  We
can now prove the following theorem about singleton sets.  We
abbreviate \((\exists x. P~x)\land (\forall x_1\forall
x_2. P~x_1\supset P~x_2\supset x_1=x_2)\) by $\exists!x.P~x$ in the
following theorem.

\begin{theorem}\label{thm:compute}
Assume that $P$ is a $\Pos_1$ expression of type $i\ra o$ and that
$\exists! y. P y$ has a \mumall proof.  There is a sequence of
transitions from the initial state $\state{y}{P~y}{y}$ to a success
state of value $t$ if and only if $P~t$ has a \mumall proof.
\end{theorem}

\begin{proof}
The forward direction is immediate: given a sequence of transitions
from the initial state $\state{y}{P~y}{y}$ to the success state
$\state{\cdot}{\cdot}{t}$, it is easy to build a \mumall proof of
$P~t$.  Conversely, assume that there is a \mumall proof of $P~t$ for
some term $t$ and, hence, of $\exists y. P~y$.  By
Lemma~\ref{lemma:compute}, there is a sequence of transitions from the
initial state $\state{y}{P~y}{y}$ to the success state
$\state{\cdot}{\cdot}{s}$, where $P~s$ has a \mumall proof.  Given a
(cut-free) \mumall proof of $\exists! y.P y$, that proof contains a
\mumall proof of $\forall x_1\forall x_2. P~x_1\supset P~x_2\supset
x_1=x_2$, which, when combined using cut with the proofs of the
formulas $P t$ and $P s$ (and the admissibility of cut for \mumall)
allows us to conclude that $t=s$.
\end{proof}

Thus, a (naive) proof-search algorithm involving both unification and
nondeter\-ministic search is sufficient for computing the functions
encoded in relations in this setting.  This result puts the
computation of such functions inside the domain of logic programming,
where relations, unification, and nondeter\-ministic proof search are
routinely encountered.  As a result, deploying any number of
Prolog-style implementation strategies, such as those found in
\cite{stickel88jar,aitkaci91wam}, can make the search for such proofs
more effective.

% LocalWords:  ministic nonde ter ly nondeter

%% file: ackermann.tex
\section{The totality of the Ackermann function}
\label{sec:ackermann}

The question of the expressivity of \mumall has been analyzed by
Baelde \cite[Section 3.5]{baelde08phd}, who provided a lower bound to
it by characterizing a subset of \mumall where proofs can be
interpreted as primitive recursive functions, and cut elimination
corresponds to computing those functions. The ideas behind the
encoding can be used in order to express primitive recursive functions
as fixed points and provide proofs of the totality of these functions in
a similar fashion to that used for the $plus$ relation.  However,
Baelde noted that this encoding is insufficient for a computational
interpretation of Ackermann's function.  Furthermore, it is also
insufficient to obtain a proof that the underlying relation represents
a total function.

We show here a different method, based on the extension to \mumallx
provided in Section~\ref{ssect:exponentials}, that allows us to prove
the totality of Ackermann's function.  The encoding of Ackermann's
function in \mumall is based on the following relational
specification.
\[
\begin{split}
       \ack = \mu(\lambda \ack\lambda m\lambda n\lambda a.~& 
(m = 0 \lltens a = s~n)\llplus\\
&\exists p(m = s~p\lltens n = 0\lltens\ack~p~(s~0)~a)\llplus \\
&\exists p\exists q\exists b(m = s~p\lltens n = s~q\lltens\ack~m~q~b\lltens \ack~p~b~a))
    \end{split}
\]
In order to prove that this three-place relation determines a total
function, we need to prove \emph{determinacy} (the first two
arguments uniquely determine the third argument) and \emph{totality}
(for every choice of the first two arguments, there exists a
value for the third argument).  The proof of determinacy, that is, of
the formula
\[
\forall x\forall y\forall a_1\forall a_2.(\ack~x~y~a_1\limp\ack~x~y~a_2\limp a_1=a_2),
\]
proceeds simply as follows: first, perform induction on $\ack~x~y~a_1$
using the rest of the context as an invariant; second, perform case
analysis on the three ways that $\ack~x~y~a_1$ is defined, and, third,
in each case use the inductive assumption to complete the proof.  We
have described an \mumall proof since neither contraction nor
weakening is needed in this proof.  (This proof outline also applies
to proving the determinacy of the $\plus$ relation in
Section~\ref{sec:examples}.)  In the rest of this section, we prove
the following formula regarding the totality of this relation.
\[
    \forall m\forall n(nat~m \limp nat~n \limp \exists a.~(\ack~m~n~a \lltens nat~a))
\]
We now illustrate a proof of this formula.%
\footnote{A formalization of this proof using the Abella theorem
prover~\cite{baelde14jfr} is available at\\
\url{https://www.lix.polytechnique.fr/Labo/Dale.Miller/papers/PA-and-muMALL/}.} 
In doing so, we will highlight the crucial use of the encoded
exponentials in \mumallx. For greater clarity, we will use
$\llimp$ as a shorthand, and we will retain the overline syntax for
negation instead of computing the explicit De Morgan duality.  The
proof begins by introducing the universal quantifiers and then
applying twice the $\llpar$ rule:
\[
  \infer=[\forall,\llpar]
         {\forall m\forall n(nat~m \limp nat~n \limp \exists a.~\ack~m~n~a)}{
          \oneside{\overline{nat~m}, \overline{nat~n}, \exists a.~\ack~m~n~a}}
\]
At this point, we need to use the coinduction rule twice, once with
$\overline{nat~n}$ and once with $\overline{nat~m}$.  The
coinvariants we introduce for these inductions will be where we
exploit the encoding of the exponentials. In the first induction, we
use as coinvariant the negation of the remaining context of the
sequent with a $\Bang$ added, that is $ \overline{\lambda m
  \Bang(\forall n ~nat~n \llimp \exists a (\ack~m~n~a \lltens
  nat~a))}$. This coinvariant needs to be contracted later in the
proof, hence the need for the exponential.  The left premise of the
$\nu$ rule is immediately verified since the coinvariant starts with a
$\Quest$ which we can derelict away, and we can conclude immediately
after by using the fact that a generalized initial rule $\oneside{
  \Gamma, \Gamma^\bot}$ is admissible in \mumall.\footnote{If $\Gamma$
is the multiset $\{B_1,\ldots,B_n\}$ then $\Gamma^\bot$ is
$\{\overline{B_1}\lltens\cdots\lltens\overline{B_n}\}$.}  The right
premise of the $\nu$ rule then yields the base and inductive steps. In
the base case, we need to prove $\vdash \Bang(\forall n ~nat~n \llimp
\exists a (\ack~\Zero~n~a \lltens nat~a))$, and we do this by
promoting away the exponential and then unfolding the base case of the
$ack$ definition. The inductive step gives us:
\[
  \infer[\Bang,\forall\llpar]{\oneside{\overline{\Bang(\forall n ~nat~n \llimp \exists a (\ack~x~n~a \lltens nat~a))},
 \Bang(\forall n ~nat~n \llimp \exists a (\ack~(s~x)~n~a \lltens nat~a))}}{
  \oneside{\overline{\Bang(\forall n ~nat~n \llimp \exists a (\ack~x~n~a \lltens nat~a))},
  \overline{nat~n}, \exists a (\ack~(s~x)~n~a \lltens nat~a)}}
\]

Since the coinvariant starts with a question mark,
we can promote the dualized coinvariant and continue the proof. We
have again exposed the dualized $\overline{nat~n}$ predicate,
over which we can perform the second induction. As before, we take the
entire sequent (abstracted over by $n$) and negate it, but this time
there is no need to add another occurrence of an exponential,
obtaining the coinvariant
\[
  \overline{\lambda k \Bang (\forall n ~nat~n \llimp \exists a
  (\ack~x~n~a \lltens nat~a)) \llimp \exists a(\ack~(s~x)~k~a \lltens nat~a)}.
\]
The left-hand premise of the $\nu$ rule is now exactly an instance of
$\oneside{\Gamma,\Gamma^\bot}$.  The base case and the inductive steps
for this second induction remain to be proved.  The base case (where
we need to prove the coinvariant for $k$ being $0$) is again proved by
a routine inspection of the definition of $ack$. The antecedent part
of the coinvariant can be used directly since it starts with $\Quest$.

The final step is the coinductive case, where we need to prove the
invariant for $(s~k)$ given the invariant for $k$: that is, we need to
prove the sequent
\[
\begin{split}
\vdash & \overline{{\Bang\,\forall x (nat~x \llimp \exists a (\ack~y~x~a \lltens nat~a))}
\llimp \exists a (\ack~(s~y)~k~a \lltens nat~a)}, \\
 &\Bang\,\forall x (nat~x \llimp \exists a(\ack~y~x~a \lltens nat~a))
 \llimp \exists a (\ack~(s~y)~(s~k)~a \lltens nat~a)
\end{split}
\]
%This is where we are to use the encoded exponential.
Introducing the second linear implication gives the dual of the
antecedent, which starts with $\Quest$ and, hence, is a contractable
copy of the coinvariant from the previous induction:
\[
  \Quest\, \overline{\forall x  (nat~x \llimp \exists a(\ack~y~x~a
    \lltens nat~a))}.
\]
The entire reason for using \mumallx to state coinvariants in this
proof is to make this contraction possible.  Now, we can decompose the
new coinvariant, a universally quantified implication, and
use the two copies we have obtained: one copy is provided to the
antecedent of the implication, and one copy is used to continue the
proof. The two premises of this occurrence of the $\nu$ rule are:
\[
\begin{split}
\vdash & {\Bang \forall x (nat~x \llimp \exists a (\ack~y~x~a \lltens nat~a))},
 \Quest\overline{\forall x (nat~x \llimp \exists a(\ack~y~x~a \lltens nat~a))} \\[4pt]
\vdash & \overline{\exists a (\ack~(s~y)~k~a \lltens nat~a)},\Quest\overline{\forall x (nat~x \llimp \exists a(\ack~y~x~a \lltens nat~a))},
  \exists a (\ack~(s~y)~(s~k)~a \lltens nat~a)
\end{split}
\]
The first one is immediately proved thanks to the fact that the
exponentials are dual. The second sequent is also easily proved by
unfolding the definition of $ack$ using its third case; the
exponential can be derelicted, and all the arising premises can be
proved without the exponentials.

Given that we have a \mumallx proof (hence, also a \mumall proof) of
the totality of the Ackermann relation, we can use the proof search
method in Section~\ref{sec:proof search} in order to actually compute
the Ackermann function.  Additionally, from the cut-elimination
theorem of \mumall, we obtain an interpretation as a computation via
proof normalization.

% LocalWords:  nat overline dualized contractable derelicted DM ack

%% file: peano.tex
\section{\mlkp contains Peano Arithmetic}
\label{sec:peano}

We now turn our attention to \mlkp: recall from
Section~\ref{sec:mumall} that this proof system is the result of
adding to the \mumall proof system both contraction $C$ and weakening
$W$ (see Figure~\ref{fig:structural}) as well as the cut rule.  The
consistency of \mumall follows immediately from its cut-elimination
theorem.  It is worth noting that adding contraction to some
consistent proof systems with weak forms of fixed points can make the
new proof system inconsistent.  For example, both Girard
\cite{girard92mail} and Schroeder-Heister
\cite{schroeder-heister93lics} describe a variant of linear logic
with unfolding fixed points that is consistent, but when contraction
is added, it becomes inconsistent.  In their case, negations are
allowed in the body of fixed point definitions.  (See also
\cite{grishin81iransm}.)  The following theorem proves that adding
contraction to \mumall does not lead to inconsistency.

\begin{theorem}
\mlkp is consistent: that is, the empty sequent is not provable.
\end{theorem}

\begin{proof}
Consider the sequent $\oneside{B_1,\ldots,B_n}$, where $n\ge 0$ and
where all the free variables of formulas in this sequent are contained
in the list of variables $x_1,\ldots,x_m$ ($m\ge 0$) all of type
$\iota$.  We say that this sequent is \emph{true} if for all
substitutions $\theta$ that send the variables $x_1,\ldots,x_m$ to
closed terms of type $\iota$ (numerals), the disjunction of the
unpolarized versions of the formula $B_1\theta,\ldots B_n\theta$ is
\emph{true} (in the standard model).  (The empty disjunction
is the same as false.)  A straightforward induction on
the structure of \mlkp proofs shows that all of the inference rules in
Figures~\ref{fig:mumall}, ~\ref{fig:three}, and~\ref{fig:structural}
are sound (meaning that when the premises are true, the conclusion is
true).  Thus, we have the following soundness result: if the sequent
$\oneside{B_1,\ldots,B_n}$ is provable in \mlkp, then that sequent is
true.  As a result, the empty sequent is not provable.
\end{proof}

We now show that Peano arithmetic is contained in \mlkp.  The terms
of Peano arithmetic are identical to the terms introduced in
Section~\ref{sec:terms} for encoding numerals.  The formulas of Peano
arithmetic are similar to unpolarized formulas except that they are built
from $=$, $\neq$, the propositional logical connectives $\land$,
$\TT$, $\lor$, $\FF$, and the two quantifiers $\hat\forall$ and
$\hat\exists$ (both of type $(i\ra o)\ra o$).  Such formulas can be
\emph{polarized} to get a polarized formula as described in
Section~\ref{ssec:polarized}.  Finally, all occurrences of
$\hat\forall$ and $\hat\exists$ are replaced by $\lambda B. \forall
x~(\overline{\nat~x} \llpar (B x))$ and $\lambda B. \exists x~(\nat~x
\lltens (B x))$, respectively.  Here, $\nat$ is an abbreviation for 
$\mu\lambda N\lambda n(n=\Zero\llplus\exists m(n=\s{m}\lltens N~m))$.

Most presentations of Peano arithmetic incorporate the addition and
multiplication of natural numbers as binary function symbols or
as three-place relations.  We will avoid introducing the
extra constructors $+$ and $\cdot$ and choose to encode addition and
multiplication as relations.  In particular, these are defined as the
fixed point expressions $\plus$ and $\mult$ given in
Section~\ref{sec:proofs}.  The relation between these two presentations
is such that the equality $x + y = w$ corresponds to $\plus \, x \, y
\, w$ and the equality $x \cdot y = w$ corresponds to $\mult \, x\,
y\,w$.  A more complex expression, such as $\forall x\forall y.~(x
\cdot s \, y = (x \cdot y + x))$, can similarly be written as either
\[
  \forall x\forall y\forall u.~\mult~x~(s\,y)~u\iimp~
      \forall v.~\mult~x~y~v\iimp \forall w. \plus~v~x~w\iimp u = w
\]
or as
\[
  \forall x\forall y\exists u.~\mult~x~(s\,y)~u\land~
      \exists v.~\mult~x~y~v\land \exists w. \plus~v~x~w\land u = w.
\]
A general approach to making such an adjustment to the syntax of
expressions using functions symbols to expressions using relations is
discussed from a proof-theoretic perspective in \cite{gerard17csl}.

Proofs in Peano arithmetic can be specified using the following six
axioms.
\[
\begin{array}{l@{\qquad}l}
  \forall x.~(s\, x) \neq \Zero &
  \forall x \forall y.~ (x + s \, x) = s (x + y)  \\
  \forall x \forall y.~(s\, x = s\, y) \supset ( x = y) &
  \forall x.~(x \cdot \Zero = \Zero)  \\
  \forall x.~(x + \Zero = x) &
  \forall x\forall y.~(x \cdot s \, y = (x \cdot y + x))
\end{array}
\]
and the axiom scheme (which we write using the predicate variable $A$)
\[
  (A\Zero\land\forall x.~(Ax\supset A\s{x}))\supset\forall x.\, Ax.
\]
We also admit the usual inference rules of modus ponens and universal
generalization.

\begin{theorem}[\mlkp contains Peano arithmetic]\label{thm:peano}
Let $Q$ be any unpolarized formula, and let $\hat Q$ be a polarized
version of $Q$.  If $Q$ is provable in Peano arithmetic then $\hat Q$
is provable in \mlkp.
\end{theorem}

\begin{proof}
It is easy to prove that $\mult$ and $plus$ describe precisely the
multiplication and addition operations on natural numbers.
As we illustrate next, the translations of the Peano Axioms can all be
proved in \mlkp. 
Since the presence of contraction and weakening in \mlkp means that
different polarizations of a formula are all equivalence in \mlkp, we
only need to consider proving a single such polarization.
The following formulas result from polarizing the translation of the
first two Peano Axioms.
\[
  \forall x.\overline{\nat~x}\llpar~(s\, x) \neq \Zero 
  \hbox{\quad and\quad}
  \forall x.\overline{\nat~x}\llpar
  \forall y.\overline{\nat~y}\llpar
  ~\overline{(s\, x = s\, y)}\llpar ( x = y).
\]
Given that induction is not needed to prove these formulas and that
the weakening rule is admissible for $\Neg_1$-formulas (as observed in
Section~\ref{sec:mumall} and proved in \cite{baelde07lpar,baelde12tocl}),
the proof of these two formulas are essentially the same as the proofs
given for their untranslated forms in Section~\ref{sec:examples}.  The
proofs of the next four axioms use induction in the usual way.
Thus, consider the final axiom---the induction scheme---and it
polarized translation
\[
\overline{\big(A \Zero \lltens \forall x.~(\overline{nat~x} \llpar 
                                          \overline{Ax} \llpar A\s{x})\big)}
  \llpar \forall x.~(\overline{nat~x} \llpar A x)
\]
An application of the $\nu$ rule to the second occurrence of
$\overline{nat~x}$ can provide an immediate proof of this axiom.
Finally, the cut rule in \mlkp allows us to encode modus ponens.
\end{proof}

% LocalWords:  nat

%% file: conserve.tex
\section{Conservativity results for linearized arithmetic}
\label{sec:la}

A well-known result in the study of arithmetic is the following.
\begin{quote}
Peano arithmetic is $\Pi_2$-conservative over Heyting arithmetic: if
Peano arithmetic proves a $\Pi_2$-formula $A$, then $A$ is already
provable in Heyting arithmetic~\cite{friedman78hol}.
\end{quote}
We present two conservativity theorems in this section that relate the
stronger logic \mlk to the weaker logic \mumall. 

The strongest set of theorems we explore in this section are based on
$\Neg_2$-formulas, and the richest sequents we consider contain only
$\Neg_1$, $\Pos_1$, and $\Neg_2$-formulas: such sequents are
called \emph{reduced}.  Note that the sequent
$\oneside{B_1,\ldots,B_n}$ ($n\ge2$) is reduced if and only if the
formula $B_1\llpar\cdots\llpar B_n$ is $\Neg_2$.

An $\Neg_2$-formula is \emph{pointed} if every occurrence of $B_1\llpar
B_2$ in it is such that either $B_1$ or $B_2$ is $\Neg_1$ and every
occurrence of $\nu B$ in it is such that $B$ is $\Neg_1$.  This notion
of pointed has been used in game-theoretic semantics for linear logic
\cite{delande10apal} (in the form of \emph{simple expressions}) and
model checking \cite{heath19jar} (where it was called
\emph{switchable}).  In the context of \mumall, focusing on a pointed
formula results in additive synthetic rules even when that formula
contains multiplicative connectives~\cite{heath19jar}.  The formulas
in $\Neg_1$ and $\Pos_1$ are pointed.  Also, if $B_0$ is in $\Pos_1$
then the formula $\forall \x. B_1\limp\cdots\limp B_n\limp B_0$
($n\ge1$) is a pointed formula if and only if the formulas
$B_1,\ldots,B_n$ are all in $\Pos_1$.  The formulas stating the
totality and determinacy of the $\plus$ relation in
Section~\ref{sec:proof search} and the totality of Ackermann's
relation in Section~\ref{sec:ackermann} are pointed formulas.  We say
that a reduced sequent $\oneside{B_1,\ldots,B_n}$ is \emph{pointed} if
$n=1$ and $B_1$ is pointed or $n\ge2$ and $B_1\llpar\cdots\llpar B_n$
is pointed.  Put in an equivalent way: the reduced sequent
$\oneside{B_1,\ldots,B_n}$ is pointed if and only if it contains at
most one pointed formula that is not $\Neg_1$.

\subsection{\mlk is conservative over \mumall for $\Pos_1$-formulas}

A \emph{positive region} is a (cut-free) \mlk proof that contains only
the inference rules $\mu\nu$, contraction, weakening, and the
introduction rules for the positive connectives: \ie, there are no
introduction rules for the negative connectives.
If $\oneside{\Gamma_1}$ and $\oneside{\Gamma_2}$ are sequents, we say
that $\oneside{\Gamma_1}$ is a \emph{subsequent of}
$\oneside{\Gamma_2}$ if $\Gamma_1$ is a sub-multiset of $\Gamma_2$.

\begin{lemma}\label{lem:positive region}
  Let $\oneside{\Gamma}$ be a reduced sequent that has a positive
  region proof.  There exists a pointed subsequent $\oneside{\Gamma'}$
  of $\oneside{\Gamma}$ such that $\oneside{\Gamma'}$ has a \mumall
  proof.
\end{lemma}

\begin{proof}
Let $\Xi$ be a positive region proof of $\oneside{\Gamma}$.  We
proceed by induction on the structure of $\Xi$.

If $\Xi$ is the $\mu\nu$ rule, then $\Gamma$ contains exactly two
occurrences of formulas that are complementary: since this sequent is
reduced, one of these formulas is positive and, hence, must be
$\Pos_1$ and the other formula is $\Neg_1$.  Therefore,
$\oneside{\Gamma}$ is pointed and has $\Xi$ as a \mumall proof.

Next, consider the case where the last inference rule of $\Xi$ is
either the following contraction or weakening (where $\Gamma$ is
$\{B\}\cup\Gamma_0$).
\[
  \infer[W]{\oneside{\Gamma_0,B}}{\oneside{\Gamma_0}}
  \qquad
  \infer[C]{\oneside{\Gamma_0,B}}{\oneside{\Gamma_0,B,B}}
\]
If $B$ is positive, the inductive assumption yields the conclusion
immediately.  If $B$ is an $\Neg_1$-formula, then the result follows
from the inductive assumption and the admissibility in \mumall of
these structural rules for $\Neg_1$-formulas.

Finally, consider the case where the last rule of $\Xi$ is an
introduction rule for one of the positive connectives $\lltens$,
$\llone$, $\llplus$, $=$, $\exists$, or $\mu$.  The most involved of
these cases is when the last inference rule of $\Xi$ introduces
$\lltens$.  Thus, $\Gamma$ can be written as $\{B_1\lltens
B_2\}\cup\Gamma_0$ (note that $B_1$ and $B_2$ are $\Pos_1$-formulas) and
$\Xi$ is of the form
\[
  \infer[\lltens,]{\oneside{\Gamma_0, B_1\lltens B_2}}
        {\deduce{\oneside{\Gamma_1,B_1}}{\Xi_1} &
         \deduce{\oneside{\Gamma_2,B_2}}{\Xi_2}}
\]
where $\Gamma_0$ equals $\Gamma_1\cup\Gamma_2$.  By the inductive
hypothesis, there are pointed subsequents $\oneside{\Delta_1}$ and
$\oneside{\Delta_2}$ of the left and right premises, respectively,
such that $\oneside{\Delta_1}$ and $\oneside{\Delta_2}$ have \mumall
proofs $\Xi'_1$ and $\Xi'_2$, respectively.  If $\Delta_1$ is a
subsequent of $\Gamma_1$, we can take $\Gamma'$ to be $\Delta_1$.
If $\Delta_2$ is a subsequent of $\Gamma_2$, we can take $\Gamma'$
to be $\Delta_2$.  The only remaining case is when $\Delta_1$ is
$\{B_1\}\cup\Delta'_1$ and $\Delta_2$ is $\{B_2\}\cup\Delta'_2$ for
multisets $\Delta'_1$ and $\Delta'_2$.  Note that $\Delta'_1$ and
$\Delta'_2$ contain only $\Neg_1$-formulas.  The sequent
$\oneside{\Delta_1,\Delta_2,B_1\lltens B_2}$ is then a pointed
subsequent of $\oneside{\Gamma}$ with the \mumall proof that results
from applying the $\lltens$ rule to $\Xi'_1$ and $\Xi'_2$.
\end{proof}

\begin{theorem}\label{thm:lkmall}
  \mlk is conservative over \mumall for $\Pos_1$-formulas.  That is,
  if $B$ is a $\Pos_1$-formula and $\oneside{B}$ has a \mlk proof then
  $\oneside{B}$ has a \mumall proof.
\end{theorem}

\begin{proof}
Let $B$ be a $\Pos_1$-formula.  If the sequent $\oneside{B}$ has
a \mlk proof, that proof must be a positive region since it contains
no negative subformulas.  Thus, by Lemma~\ref{lem:positive region}, we
know that that sequent contains a pointed sequent that is provable
in \mumall proof.  However, the only such sequent is $\oneside{B}$.
\end{proof}

\subsection{\mlkNeg is conservative over \mumall for $\Neg_1$-formulas}

Our next conservativity result requires restricting the complexity of
coinvariants used in the $\nu$ rule.  We say that a sequent has a
\mlkNeg proof if it has a \mlk proof in which all coinvariants of
the proof are $\Neg_1$.  This restriction on proofs is similar to the
restriction that yields the $I\Sigma_1$ fragment of Peano
Arithmetic~\cite{paris78lc}.
Note that the proof of the totality of the Ackermann function discussed in
Section~\ref{sec:ackermann} is an example of a proof that is not in \mlkNeg
since it uses complex coinvariants (involving the encoding of an exponential)
that are much richer than $\Neg_1$.

\begin{lemma}\label{lem:pointed}
If the conclusion of a \mlkNeg inference rule that introduces a
negative connective is a pointed sequent then all of the premises of
that rule are pointed.
\end{lemma}

\begin{proof}
We illustrate this proof for three of the rules introducing negative
connectives.  The remaining cases are similar and simpler.

Assume that in the following inference rule, the sequent
$\oneside{\Gamma, B\llpar C}$ is pointed. 
\[
  \infer[\llpar]{\oneside{\Gamma, B\llpar C}}{\oneside{\Gamma, B, C}}
\]
We have three cases to consider.  $(i)$ All members of
$\Gamma\cup\{B\llpar C\}$ are $\Neg_1$-formulas.  In that case, the
premise contains only $\Neg_1$-formulas and is, thus, pointed.
$(ii)$ $B\llpar C$ is not $\Neg_1$.  Thus, it is $\Neg_2$.  Since it
is a pointed formula, either $B$ is $\Neg_1$ and $C$ is $\Neg_2$ or
$C$ is $\Neg_1$ and $B$ is $\Neg_2$.  In either case, the premise is
pointed.  $(iii)$ There is a formula in $\Gamma$ that is not $\Neg_1$.
Thus, $B\llpar C$ is $\Neg_1$ and so are $B$ and $C$.  The premise is
again pointed.

Assume that in the following inference rule, the sequent
$\oneside{\Gamma, B\llwith C}$ is pointed. 
\[
  \infer[\llwith]{\oneside{\Gamma, B \llwith C}}
                 {\oneside{\Gamma, B} & \oneside{\Gamma, C}}
\]
We again have three cases to consider.  $(i)$ All members of
$\Gamma\cup\{B\with C\}$ are $\Neg_1$-formulas.  In that case, the
premises contain only $\Neg_1$-formulas and are, thus, pointed.
$(ii)$ $B\llpar C$ is not $\Neg_1$.  Thus, it is $\Neg_2$.  Since it
is a pointed formula, either $B$ is $\Neg_1$ and $C$ is $\Neg_2$ or
$C$ is $\Neg_1$ and $B$ is $\Neg_2$.  In either case, the premises are 
pointed.  $(iii)$ There is a formula in $\Gamma$ that is not $\Neg_1$.
Thus, $B\llwith C$ is $\Neg_1$ and so are $B$ and $C$.  The premises
are again pointed.

Assume that in the following inference rule, the sequent
$\oneside{\Gamma, \nu B\t}$ is pointed. 
\[
  \infer[\nu]{\oneside{\Gamma, \nu B\t}}
             {\oneside{\Gamma, S\t} & \oneside{BS\x, \overline{(S\x)}}}
\]
Since we are in \mlkNeg, both $B$ and $S$ are in $\Neg_1$.  Hence, the
right premise is pointed.  The left premise is also pointed since both
$S\t$ and $\nu B\t$ are $\Neg_1$-formulas.
\end{proof}

\begin{theorem}
  \mlkNeg is conservative over \mumall for $\Neg_1$-formulas.  That
  is, if $B$ is an $\Neg_1$-formula and $\oneside{B}$ has a \mlk proof,
  then $\oneside{B}$ has a \mumall proof.
\end{theorem}

\begin{proof}
Let $\Gamma$ be a multiset of $\Neg_1$-formulas and let $\Xi$ be an
\mlkNeg proof of $\Gamma$.  Note that all occurrences of formulas in
all sequents in $\Xi$ are in $\Neg_1$.  We proceed by induction on the
number of structural rules (weakening and contraction) that occur in
$\Xi$.  If this number is zero, then $\Xi$ is the desired \mumall
proof.  Otherwise, assume that there is a structural rule and choose
one uppermost occurrence.  For example, if this structural
rule is the contraction
\[
  \infer[,]{\oneside{N,\Delta}}{\oneside{N,N,\Delta}}
\]
then the premise has a \mumall proof.  By the admissibility of
contraction in \mumall for $\Neg_1$-formulas \cite[Proposition
  2.12]{baelde12tocl}, we know that $\oneside{N,\Delta}$ has a \mumall
proof.  A similar argument holds if the uppermost structural rule is
weakening since weakening is similarly admissible for $\Neg_1$
formulas in \mumall.  As a result, we can build a new \mlkNeg proof
$\Xi'$ where this uppermost structural rule is replaced with a \mumall
proof, thus reducing the number of structural rules from those in $\Xi$.
\end{proof}

\subsection{\mlkNeg is conservative over \mumall for pointed formulas}

We can conclude from the two preceding theorems that \mlkNeg is
conservative over \mumall for both $\Pos_1$ and $\Neg_1$-formulas.  We
can generalize these two conservativity results to $\Neg_2$-formulas
if we restrict such formulas to be pointed.

\begin{theorem}\label{thm:lk1mall}
  \mlkNeg is conservative over \mumall for pointed
  $\Neg_2$-formulas.  That is, if $B$ is a pointed formula
  such that $\oneside{B}$ has a \mlkNeg proof, then $\oneside{B}$
  has a \mumall proof.  
\end{theorem}

The proof of this theorem would be greatly aided if we had a focusing
theorem for \mlk.  If we take the focused proof system for \mumall
given in \cite{baelde12tocl,baelde07lpar} and add contraction (in the
decide rule) and weakening (in the initial rule), we have a natural
candidate for a focused proof system for \mlk.  (The focused proof
systems \LKF and \MALLF in \cite{liang24psh} have exactly these
differences.)  However, the completeness of that focused proof system is
currently open.  As Girard points out in~\cite{girard91mscs}, the
completeness of such a focused (cut-free) proof system would allow the
extraction of the constructive content of classical $\Pi^0_2$
theorems, and we should not expect such a result to follow from the
usual ways that we prove cut elimination and the completeness of
focusing.  As a result of not possessing such a focused proof system
for \mlk, we must now reproduce aspects of focusing to prove
Theorem~\ref{thm:lk1mall}.  We shall return to prove this theorem at
the end of this section.

\begin{lemma}
Let $\oneside{\Gamma}$ be a sequent containing only pointed
formulas.  Every formula occurrence in a sequent occurring in a \mlkNeg
proof of $\oneside{\Gamma}$ is pointed.
\end{lemma}

\begin{proof}
Let $\oneside{\Gamma}$ be a sequent in which every formula is
pointed, and let $\Xi$ be a \mlkNeg proof of this sequent.  We now
prove by induction on the structure of $\Xi$ that every formula
occurrence in every sequent in that proof is pointed.  Since
subformulas of a pointed formula are pointed, the only inference
rules we need to check for this property explicitly are the rules
for fixed points since they do not obey the subformula property.  In
particular, consider the derivation
\[
  \infer[\nu\qquad\hbox{and}\qquad]
        {\oneside{\Gamma, \nu B\t}}
             {\oneside{\Gamma, S\t} & \oneside{BS\x, \overline{(S\x)}}}
  \infer[\mu.]{\oneside{\Gamma, \mu B\t}}{\oneside{\Gamma, B(\mu B)\t}}
\]
In the $\nu$ rule, both $B$ and $S$ are $\Neg_1$  and, hence,
$B S\x$ is $\Neg_1$ and $\overline{(S\x)}$ is $\Pos_1$.  As a result,
the right premise of the $\nu$ rule contains only pointed
formulas.  The inductive assumption yields the same conclusion for the
left premise.  In the $\mu$ rule, the occurrence of $\mu B\t$ is
$\Pos_1$ hence so is $B(\mu B)\t$.  This case follows from the
inductive hypothesis.
\end{proof}

Let $\oneside{P,N,\Gamma}$ be a sequent where $P$ is a positive
formula, and $N$ is a negative formula.  Assume that we have a proof of
this sequent where the $P$ formula is introduced at the root and the
$N$ formula is introduced on one of the premises of that rule.  As is
known from \mumall (see \cite{baelde07lpar,baelde12tocl}), all
occurrences of a positive connective introduced
immediately below the introduction of a negative connective
can be permuted so that the negative connectives are introduced
immediately below the positive connective.  As a result, we shall
introduce the following normal forms of proofs: an \mlk proof is
a \pnproof if it is a \mlkNeg proof in which there is no occurrence of a
negative introduction rule above a positive introduction rule.  By
permuting inference rules in (cut-free) proofs, it is easy to prove
the following lemma.

\begin{lemma}\label{lem:pnproof}
If a sequent has an \mlkNeg proof, it was a \pnproof.
\end{lemma}

Given the structure of \pnproofs, the following lemma has a direct
proof.

\begin{lemma}\label{lem:posreg}
If a reduced sequent has a \mlkNeg proof in which the last inference
rule is the introduction of a positive connective, then it has a
positive region proof.
\end{lemma}

Ketonen \cite{ketonen22book} used cut admissibility in Gentzen's \LK
proof system to provide elegant proofs of the invertibility for some
introduction rules.  His arguments can be used to show the
invertibility of the introduction rules for the (negative) connectives
$\bot$, $\llpar$, $\top$, $\with$, and $\forall$ in \mumall.  Since we
do not have a cut-admissibility result for \mlk, we will need to make
a simple variation on his proof to prove similar statements for \mlk.

\begin{lemma}\label{lem:invert}
Let $\oneside{N,\Gamma}$ be a sequent where the top-level connective of $N$ 
is one of the following (negative) connectives:
$\neq$, $\bot$, $\llpar$, $\top$, $\with$, and $\forall$.  If this
sequent has a \mlk proof, it has a \mlk proof in which the last
inference rule introduces $N$.
\end{lemma}

\begin{proof}
\newcommand{\fragmentA}{
   \infer[\oplus]{\oneside{B_1, \overline{B_1\with B_2}}}{
         \infer[\init]{\oneside{B_1, \overline{B_1}}}{}}}
\newcommand{\fragmentB}{
   \infer[\lltens]{\oneside{B_1,B_2,\overline{B_1\llpar B_2}}}
         { \infer[\init]{\oneside{B_1, \overline{B_1}}}{} &
           \infer[\init]{\oneside{B_2, \overline{B_2}}}{}}}
Assume that the $\oneside{B_1\with B_2,\Gamma}$ sequent has a \mlk
proof $\Xi$.  We want to prove that it has a \mlk proof in which the
last inference rule introduces $B_1\with B_2$.  Consider the
following \mlkp proof.
\[
\infer[\with]{\oneside{\Gamma, B_1\with B_2}}{
  \infer[\cut]{\oneside{\Gamma, B_1}}
        {\deduce{\oneside{\Gamma, B_1\with B_2}}{\strut\Xi} & 
         \fragmentA} &
  \infer[\cut]{\oneside{\Gamma, B_2}}
        {\deduce{\oneside{\Gamma, B_1\with B_2}}{\strut\Xi} & 
         \fragmentA}}
\]
Since \init is admissible in \mumall, it is admissible in \mlk.  If we
can eliminate the \cut rules from this proof, we will have a \mlk
proof in which the last inference rule introduces $B_1\with B_2$.  We
can move the \cut rule upwards to eliminate it
in the usual fashion.  In this case, the only issue arrives when the
last inference rule of $\Xi$ is a contraction on $B_1\with B_2$.
That is, the left premises of the $\with$-introduction rule above is
of the form
\[
  \infer[\cut.]{\oneside{\Gamma, B_1}}
        {\infer[C]{\oneside{\Gamma, B_1\with B_2}}
                  {\deduce{\oneside{\Gamma, B_1\with B_2, B_1\with B_2}}
                          {\strut\Xi'}} & 
         \fragmentA}
\]
This derivation can be transformed into the following derivation.
\[
  \infer[C]{\oneside{\Gamma, B_1}}{
  \infer[\cut]{\oneside{\Gamma, B_1,B_1}}
        {\infer[\cut]{\oneside{\Gamma, B_1, B_1\with B_2}}
                     {\deduce{\oneside{\Gamma, B_1\with B_2, B_1\with B_2}}
                             {\strut\Xi'} & \fragmentA
} & 
         \fragmentA}}
\]
This way, the contraction applied to $B_1\with B_2$ is
transformed into contractions on the subformulas $B_1$ and $B_2$.
If we replace contraction with weakening, a similar transformation can
be done.
\[
  \infer[\cut]{\oneside{\Gamma, B_1}}
        {\infer[W]{\oneside{\Gamma, B_1\with B_2}}
                  {\deduce{\oneside{\Gamma}}
                          {\strut\Xi'}} & 
         \fragmentA}
         \qquad\Longrightarrow
  \infer[W]{\oneside{\Gamma, B_1}}{\deduce{\oneside{\Gamma}}{\strut\Xi'}}
\]
In this way, the weakening applied to $B_1\with B_2$ is
transformed into weakenings on the subformulas $B_1$ and $B_2$.

Similarly, assume that the $\oneside{B_1\llpar B_2,\Gamma}$ sequent has
a \mlk proof $\Xi$ and that $\Xi$ ends in a contraction.
\[
  \infer[\llpar]{\oneside{\Gamma, B_1\llpar B_2}}{
    \infer[\cut]{\oneside{\Gamma, B_1,B_2}}
          {\infer[C]{\oneside{\Gamma, B_1\llpar B_2}}
                    {\deduce{\Gamma, B_1\llpar B_2, B_1\llpar B_2}
                            {\strut\Xi'}}
         & \fragmentB}}
\]
This derivation can be transformed into the following derivation in
which contraction and cut have been switched.
\[
  \infer[\llpar]{\oneside{\Gamma, B_1\llpar B_2}}{
  \infer=[C\times 2]{\oneside{\Gamma, B_1,B_2}}{
  \infer[\cut]{\oneside{\Gamma, B_1,B_1, B_2,B_2}}
              {\infer[\cut]{\Gamma, B_1, B_2, B_1\llpar B_2}
                     {\deduce{\Gamma, B_1\llpar B_2, B_1\llpar B_2}
                             {\strut\Xi'}
                      &\fragmentB}
     & \fragmentB}}}
\]
Again, the contraction applied to $B_1\llpar B_2$ is
transformed into contractions on the subformulas $B_1$ and $B_2$.  

The remaining cases involving $\neq$, $\bot$, $\top$, and $\forall$
are similar and simpler.  In this way, we prove the invertibility of
the introduction for all negative connectives except those for $\nu$.
\end{proof}

Note that the invertibility of these negative connectives implies that the
contraction and weakening rule does not need to be applied to formulas
with such negative connectives at their top level.  In this way,
Lemma~\ref{lem:invert} allows us to replace contractions on negative
formulas by contractions on their subformulas that are either
$\nu$-expressions (which we shall deal with next) or on their positive
subformulas.

For the purposes of the rest of this section, we generalize the $\nu$
rule to incorporate instances of the structural rules applied to a
$\nu$-formula.
\[
  \infer[\Cnu{n}\quad n\ge0]{\oneside{\Gamma,\nu B \t}}
        {\oneside{\Gamma,S_1\t, \ldots, S_n\t}\quad
         \oneside{B S_1 \x,\overline{S_1 \x}}
         \quad\cdots\quad
         \oneside{B S_n \x,\overline{S_n \x}}}
\]
Since we are working within \mlkNeg, the coinvariants $S_1,\ldots,S_n$
are $\Neg_1$.  This rule has $n+1$ premises.  If $n=0$ this rule is
the weakening rule for $\nu$-expression; if $n=1$ this rule is the
$\nu$ rule; and if $n\ge 2$, it can be justified using $n-1$
contractions.  For example, when $n=2$ the following combination of $\nu$
and contraction rules justify the inference rule $\Cnu{2}$.
\[
  \infer[C]{\oneside{\Gamma,\nu B \t}}{
  \infer[\nu]{\oneside{\Gamma,\nu B \t,\nu B \t}}{
  \infer[\nu]{\oneside{\Gamma,\nu B \t,S\t}}
             {\oneside{\Gamma,S\t, U\t}\quad
              \oneside{B U \x,\overline{U \x}}}
      &
      \oneside{B S \x,\overline{S \x}}}}
\]

Let \mlkNegs be the proof system that results from formally removing
the $\nu$ rule and adding the rule $\Cnu{i}$ for every natural number
$i$.  Every \mlkNeg proof can be converted to a \mlkNegs
proof by simply renaming the $\nu$ rule as $\Cnu{1}$.

Given that the side formulas $\Gamma$ in the $\Cnu{}$ rule appear in
only one premise, it is easy to show that this rule permute down over
any introduction rule for positive connectives.  For this reason, we
shall assume that the definition (and completeness) of \pnproofs 
are extended to include the $\Cnu{}$ rule as a negative rule.

\begin{lemma}\label{lem:no neg contraction}
If $B$ is an $\Neg_2$-formula with a \mlkNegs proof then it has
a \mlkNegs proof in which the contraction rule is used only with
positive formulas.
\end{lemma}

\begin{proof}
As Lemma~\ref{lem:invert} shows, if one of the negative connectives
other than $\nu$ appears in a sequent, we can assume that that
connective is immediately applied: in particular, a contraction is not
used.  As the proof of Lemma~\ref{lem:invert} illustrated,
contractions on negative formulas can be permuted to be contractions
on positive subformulas or integrated into the $\Cnu{}$ rule.
\end{proof}

We now return to Theorem~\ref{thm:lk1mall} and provide it with a
proof.  Let $B$ be a pointed formula such that
$\oneside{B}$ has a \mlkNegs proof.  We need to show that
$\oneside{B}$ has a \mumall proof, and we do that by transforming the
\mlkNegs proof into a proof without contractions and weakenings.  We
have shown that we can eliminate the use of these two rules with
negative formulas, and when they are used with positive formulas, we
can move them into a positive region where we know their use is
superfluous.  More specifically, assume that there is a \mlkNegs proof
of $\oneside{B}$ that is a \pnproof (by Lemma~\ref{lem:pnproof}) in
which contraction and weakening are not applied to negative formulas
(by Lemma~\ref{lem:no neg contraction}).  Let $\Xi$ be a \mlkNegs
proof satisfying these restrictions.  Since the conclusion of $\Xi$,
namely $\oneside{B}$, is a pointed sequent, Lemma~\ref{lem:pointed}
implies that the premise and conclusion of all introductions of
negative connectives are pointed.  Now consider a (pointed) sequent
$\oneside{\Gamma}$ that is the premise of the introduction of a
negative connective while also being the conclusion of a positive
region proof.  By Lemma~\ref{lem:positive region}, we know that there
is a pointed subsequent $\oneside{\Gamma'}$ of $\oneside{\Gamma}$ that
has a \mumall proof.  The only difference between $\Gamma$ and
$\Gamma'$ is that the former can have $\Neg_1$-formulas, not in the
latter.  We can then take the \mumall proof for $\oneside{\Gamma'}$
and use the admissibility in \mumall of weakening for $\Neg_1$
formulas to provide a \mumall proof of $\oneside{\Gamma}$.  The only
remaining detail is to prove that the instances of $\Cnu{}$ used in
\mlkNegs proofs are admissible in \mumall: this is easy to show by
using the admissibility in \mumall of the weakening and contraction of
$\Neg_1$-formulas.

% LocalWords:  subsequents

%% file: related.tex
\section{Related and future work}
\label{sec:related}
\vspace{-0.5ex}

% Work on the arithmetic hierarchy

The main difference between the hierarchy of $\Pos_n/\Neg_n$-formulas
used in this paper and the familiar classes of formulas in the
arithmetic hierarchy (based on quantifier alternations) is the
occurrences of fixed points within formulas.  In that regard, \mumall
and \mlk are probably more aptly compared to the extension of Peano
Arithmetic based on general inductive definitions found in
\cite{mollerfeld02phd}.

% Other work on mumall - cyclic

Circular proof systems for logics with fixed points have received much
attention in recent years, especially within the context of
linear logic \cite{baelde22lics,de22fsttcs,ehrhard21lics}
and intuitionistic logic~\cite{curzi23lics}.  Such proof systems
generally eschew all first-order term structures (along with
first-order quantification).  They also eschew the use of explicit
invariants and use cycles within proofs as an implicit
approach to discovering invariants.

% Automation related to mumall - Bedwyr, Baelde/Miller/Snow, Abella

Historically speaking, the logic \mumall was developed along with the
construction of the Bedwyr model checker~\cite{baelde07cade}.
Although that model checker was designed to prove judgments in
classical logic, it became clear that only linear logic principles
were needed to describe most of its behaviors.  The paper
\cite{heath19jar} illustrates how \mumall and its (partial)
implementation in Bedwyr can be used to determine standard
model-checking problems such as reachability and simulation.  A small
theorem-proving implementation based on the focused proof system
\mumall is described in \cite{baelde10ijcar}: that prover was capable
of proving automatically many of the theorems related to establishing
determinacy and totality of $\Pos_1$ relational specifications.

As mentioned above, whether or not \mlk satisfies a cut-elimination
theorem or has a (relatively) complete focused proof system are open
questions.  Resolving both of these questions is an 
important research problem to consider next.

When we know that the rules for contraction and induction are not
involved (as in Section~\ref{sec:proof search}), then proof search in
\mumall resembles computation in the logic programming setting (\ie,
involving unification and nondeterministic search). In \mumall (in
contrast to \mlk), contraction is not available, leaving the
generation of coinvariants as the key feature to concentrate on for
automation.  A potentially valuable application of our work on \mumall
is to structure a theorem prover so that the cleverness involved with
discovering induction coinvariants could instead be placed on the
discovery of lemmas.  In particular, one could always choose to use
the ``obvious coinvariant'' when attempting a coinductive proof, much
as was done in the prover described in \cite{baelde10ijcar} (also,
reminiscent of the Boyer-Moore theorem prover~\cite{boyer79}).  The
cleverness required to complete the proofs could be transferred to
discovering applicable lemmas.  If appropriately organized, such a
proof would only require the user to supply a sequence of lemmas; all
the remaining details, such as case analysis and coinvariant
generation, would be automated.

Many of the results in this paper are based on Chapter 3 of the first
author's Ph.D. dissertation~\cite{manighetti22phd} and the technical
report~\cite{manighetti23arxiv}.

\vspace{-1.5ex}
\section{Conclusions}
\label{sec:conc}
\vspace{-0.5ex}

In this paper, we have started exploring \mumall as a linearized
version of arithmetic in a way similar to using Heyting
Arithmetic as a constructive version of arithmetic.  In particular, we
have considered three different proof systems.  The first is \mumall,
for which a cut-admissibility theorem is known.  The other two are
natural variants of \mumall that introduce into \mumall the rules of
contraction and weakening, yielding \mlk, as well as cut, yielding
\mlkp.  We demonstrate that the third proof system is consistent and
powerful enough to encompass all Peano Arithmetic.  While it is
known that \mumall can prove the totality of primitive recursive
function specifications, we demonstrate that the non-primitive
recursive Ackermann function can also be proved total in \mumall.  We
have also demonstrated that if we can prove in \mlk that a certain
$\Pos_1$ relational specification defines a function; then a simple
proof search algorithm can compute that function using unification and
backtracking search.  This approach differs from the proof-as-program
interpretation of a constructive proof of the totality of a relational
specification.  We have also shown a few simple cases when \mlk is
conservative over \mumall.

\smallskip\noindent\textbf{Acknowledgment:} We thank the anonymous
reviewers of an earlier draft of this paper for their valuable
comments.  We also thank Anupam Das for several conversations and
suggestions that helped shape this paper's overall structure and
results.

%% file: bib.tex
\newcommand{\Plato}[1]{}